# Robust Multi-biometric Recognition Using Face and Ear Images

Nazmeen Bibi Boodoo*, R K Subramanian
Computer Science Department
University of Mauritius
Mauritius
nazmeen182@yahoo.com

*Abstract*: **This study investigates the use of ear as a biometric for authentication and shows experimental results obtained on a newly created dataset of 420 images. Images are passed to a quality module in order to reduce False Rejection Rate. The Principal Component Analysis ("eigen ear") approach was used, obtaining 90.7 % recognition rate. Improvement in recognition results is obtained when ear biometric is fused with face biometric. The fusion is done at decision level, achieving a recognition rate of 96 %.**

I. INTRODUCTION

Ear recognition has received considerably less attention than many alternative biometrics, including face, fingerprint and iris recognition. Ear-based recognition is of particular interest because it is non-invasive, and because it is not affected by environmental factors such as mood, health, and clothing [11]. Also, the appearance of the auricle (outer ear) is relatively unaffected by aging, making it better suited for long-term identification.

Ear images can be easily taken from a distance without knowledge of the person concerned. Therefore ear biometric is suitable of surveillance, security, access control and monitoring applications. Earprints, found on the crime scene, have been used as a proof in over few hundreds cases in the Netherlands and the United States [14]. The purpose of the proposed paper is to investigate whether the integration of face and ear biometrics can achieve higher performance that may not be possible using a single biometric indicator alone.

II. EAR BIOMETRIC

Two studies performed by Iannarelli [2] provide enough evidence to show that ears are unique biometric traits. The first study compared over 10,000 ears drawn from a randomly selected sample in California, and the second study examined fraternal and identical twins, in which physiological features are known to be similar. The evidence from these studies supports the hypothesis that the ear contains unique physiological features, since in both studies all examined ears were found to be unique though identical twins were found to have similar, but not identical, ear structures especially in the Concha and lobe areas. Fig 1 shows the anatomy of the ear [3].

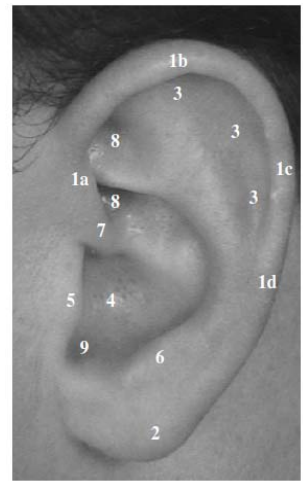

Figure 1.   1 Helix Rim, 2 Lobule, 3 Antihelix, 4 Concha, 5 Tragus, 6 Antitragus, 7 Crus of Helix, 8 Triangular Fossa, 9 Incisure Intertragica

The medical literature reports [2] that ear growth after the first four months of birth is proportional. It turns out that even though ear growth is proportional, gravity can cause the ear to undergo stretching in the vertical direction. The effect of this stretching is most pronounced in the lobe of the ear, and measurements show that the change is non-linear. The rate of stretching is approximately five times greater than normal during the period from four months to the age of eight, after which it is constant until around 70 when it again increases.

The main drawback of ear biometrics is that they are not usable when the ear of the subject is covered [2]. In the case of active identification systems, this is not a drawback as the subject can pull his hair back and proceed with the authentication process. The problem arises during passive identification as in this case no assistance on the part of the subject can be assumed. In the case of the ear being only partially occluded by hair, it is possible to recognize the hair and segment it out of the image.

III. RELATED WORK

Several Studies have been done in using ear as a biometric. The following sections give an overview of previous works done.

*A. Ear Biometric*

One of the earliest ear detection methods uses Canny edge maps to detect the ear contour [3]. Chang et al. [12] compared ear recognition with face recognition using a standard principal components analysis (PCA) technique. Recognition rate obtained were 71.6 % and 70.5 % for ear and face recognition respectively. Hurley et al. [13] considered a "force field" feature extraction approach that is based on simulated potential energy fields. They reported improved performance over PCA-based methods.

Alvarez et al. [1] used a modified active contour algorithm and Ovoid model for detecting the ear. Yan and Bowyer [8] proposed taking a predefined sector from the nose tip to locate the ear region. The non-ear portion from that sector is cropped out by skin detection and the ear pit was detected using Gaussian smoothing and curvature estimation. Then, they applied an active contour algorithm to extract the ear contour. The system is automatic but fails if the ear pit is not visible.

Li Yuan and Mu [9] used a modified CAMSHIFT algorithm to roughly track the profile image as the region of interest (ROI). Then, contour fitting is operated on ROI for further accurate localization using the contour information of the ear. Saleh et al. [18] tested a dataset of ear images using several image-based classifiers and feature-extraction methods. Classification accuracy ranged from 76.5% to 94.1% in the experiments.

Most recently, Islam et al. [5] proposed an ear detection approach based on the AdaBoost algorithm [7]. The system was trained with rectangular Haar-like features and using a dataset of varied races, sexes, appearances, orientations and illuminations. The data was collected by cropping and synthesizing from several face image databases. The approach is fully automatic, provides 100% detection while tested with 203 non-occluded images and also works well with some occluded and degraded images.

As summarized in the survey of Pun et al. [6] most of the proposed ear recognition approaches use either PCA (Principal Component Analysis) or the ICP algorithm for matching. Choras [4] proposed a different automated geometrical method. Testing with 240 images (20 different views) of 12 subjects, 100% recognition rate is reported.

The first ever ear recognition system tested with a larger database of 415 subjects is proposed by Yan and Bowyer [8]. Using a modified version of the ICP, they achieved an accuracy of 95.7% with occlusion and 97.8 % without occlusion (with an Equal-error rate (EER) of 1.2%). The system does not work well if the ear pit is not visible.

Islam et al. [10] proposed a method for cropping 3D profile face data for ear detection and applied the Iterative Closest Point (ICP) algorithm for recognition of the ear at different mesh resolutions of the extracted 3D ear data. The system obtains a recognition rate of 93%. It is fully automatic and does not rely on the presence of a particular feature of the ear (e.g. ear pit).

*B. Face Biometric*

Research in automatic face recognition dates back at 1960's [19]. A survey of face recognition techniques has been given by Zhao et al., (2003). In general, face recognition techniques can be divided into two groups based on the face representation they use:
1. Appearance-based: which uses holistic texture features and is applied to either whole-face or specific regions in a face image;
2. Feature-based: which uses geometric facial features (mouth, eyes, brows, cheeks etc.) and geometric relationships between them.

Kirby and Sirovich were among the first to apply principal component analysis (PCA) to face images, and showed that PCA is an optimal compression scheme that minimizes the mean squared error between the original images and their reconstructions for any given level of compression [20]. Turk and Pentland popularized the use of PCA for face recognition [21]. They used PCA to compute a set of subspace basis vectors (which they called "eigenfaces") for a database of face images, and projected the images in the database into the compressed subspace. New test images were then matched to images in the database by projecting them onto the basis vectors and finding the nearest compressed image in the subspace (eigenspace).

Researchers began to search for other subspaces that might improve performance. One alternative is Fisher's linear discriminant analysis (LDA, a.k.a. "fisherfaces") [22]. For any N-class classification problem, the goal of LDA is to find the N-1 basis vectors that maximize the interclass distances while minimizing the intra-class distances. At one level, PCA and LDA are very different: LDA is a supervised learning technique that relies on class labels, whereas PCA is an unsupervised technique.

One characteristic of both PCA and LDA is that they produce spatially global feature vectors. In other words, the basis vectors produced by PCA and LDA are non-zero for almost all dimensions, implying that a change to a single input pixel will alter every dimension of its subspace projection. There is also a lot of interest in techniques that create spatially localized feature vectors, in the hopes that they might be less susceptible to occlusion and would implement recognition by parts. The most common method for generating spatially localized features is to apply independent component analysis (ICA) to produce basis vectors that are statistically independent [23].

### C. Ear Versus Face Biometric

Though face recognition has been extensively studied in the past decades, imaging problems (e.g., lighting, shadows, scale, and translation) make it difficult to build an unconstrained face Identification. Also, it is difficult to collect consistent features from the face as it is arguably the most changing part of the body due to facial expressions, cosmetics, facial hair and hair styling [3]. The combination of the typical imaging problems of feature extraction in an unconstrained environment, and the changeability of the face, explains the difficulty of automating face biometrics.

Colour distribution is more uniform in ear than in human face. Not much information is lost while working with grayscale or binarised images. Ear is also smaller than face, which means that it is possible to work faster and more efficiently with images with the lower resolution. Ear images cannot be disturbed by glasses, beard or make-up. However, occlusion by hair and earring is possible.

### D. Multi-Biometric

Although most biometric systems deployed in real-world applications are unimodal, so they rely on the evidence of a single source of information for authentication, these systems have to contend with a variety of problems such as noise in sensed data, intra-class variations, inter-class similarities, non-universality, and spoof attacks. Some of the limitations imposed by unimodal biometric systems can be overcome by including multiple sources of information for establishing identity. These systems allow the integration of two or more types of biometric systems. Integrating multiple modalities in user verification and identification leads to high performance [17].

## IV. METHODOLOGY

### A. Dataset

A multimodal dataset was created. It involves people aged from 20 to 50 years old. The Kodak digital camera of 7.1 Mega pixels was used. 30 persons were involved, each one having 7 face images and 7 ear images, giving a total of 420images. To obtain ear images, the profile images were taken and cropped. Face images are of 150 × 200 resolution while ear images are of 100 × 150 resolution. The setup for the image capture is shown in Fig 2. Example of the dataset is given Fig 3 and Fig 4.

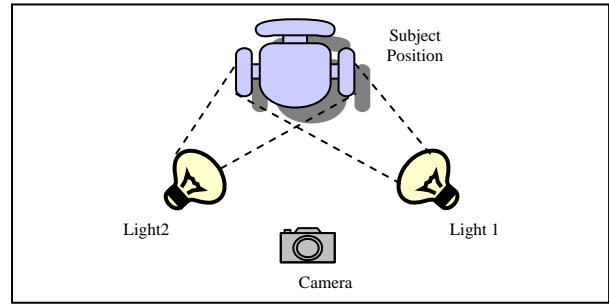

Figure 2. Image Capture Setup

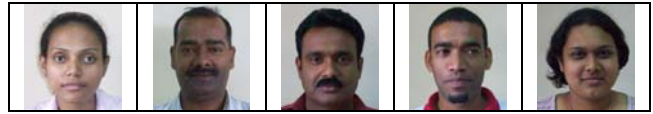

Figure 3. Sample Face images

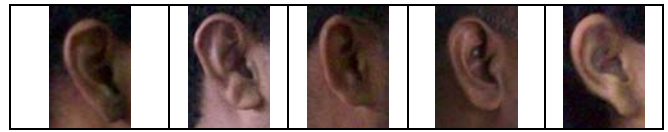

Figure 4. Sample Ear Images

The ear images have been manually cropped and resized from the original profile head images.

### B. Image Quality

The quality of biometric sample has significant impact on performance of recognition. One of the main reasons for matching errors in biometric systems is poor-quality images. Automatic biometric image quality assessment may help improve system performance.

In this study, Normalised Cross-correlation is used as a measure to determine the quality of an input image. The basis of using correlation as a pattern matching method lies in determining the degree to which the object under examination resembles that contained in a given reference image. The degree of resemblance is a simple statistic on which to base decisions about the object [25]. The so called normalised cross-correlation method is a widely used match measure in correlation based pattern recognition. For input image f and mean image in of training set, g, the normalised cross-correlation measure of match is defined as

$$\Sigma(f,g) = \frac{\Sigma_{(i,j)} f_{ij} \cdot g_{ij}}{\sqrt{\Sigma_{(i,j)} f_{ij}^2} \cdot \sqrt{\Sigma_{(i,j)} g_{ij}^2}} \quad (1)$$

## C. Face and Ear Verification

The features extracted were based on the Karhunen-Loeve (KL) expansion, also known as principal component analysis (PCA). The main reasons to used KL expansion were that it has been exhaustively studied and have proved to be quite invariant and robust when proper normalization is applied over the faces [15]. On the other hand, the main disadvantages of KL methods are its complexity and that the extracted base is data-dependent: if new images are added to the database the KL base need to be recomputed. The main idea is to decompose a face picture as a weighted combination of the orthonormal base provided by the KL transform. The base corresponds to the eigenvectors of the covariance matrix of the data, known as eigenfaces or eigenears.

Thus, the decomposition of a face image into an eigenface space provides a set of features. The maximum number of features is restricted to the number of images used to compute the KL transform, although usually only the more relevant features are selected, removing the ones associated with the smallest eigenvalues. In the classic eigenface method, proposed by Turk and Pentland [16], the PCA is performed on a dataset of face images from all users to be recognized.

## D. Levels of Fusion

Because of the use of multiple modalities, fusion techniques should be established for combining the different modalities. Integration of information in a Multimodal biometric system can occur in three main levels, namely feature level, matching level or decision level [18]. At feature level, the feature sets of different modalities are combined. Fusion at this level provides the highest flexibility but classification problems may arise due to the large dimension of the combined feature vectors. Fusion at matching level is the most common one, whereby the scores of the classifiers are usually normalized and then they are combined in a consistent manner. At fusion on decision level each subsystem determines its own authentication decision and all individual results are combined to a common decision of the fusion system.

In this study, fusion at the decision level is applied for data fusion of the various modalities, based on the majority vote rule. For three samples, as is the case, a minimum of two accept votes is needed for acceptance. Also, for the final fusion, the AND rule is used. Fig 5 shows two-level fusion applied in this study.

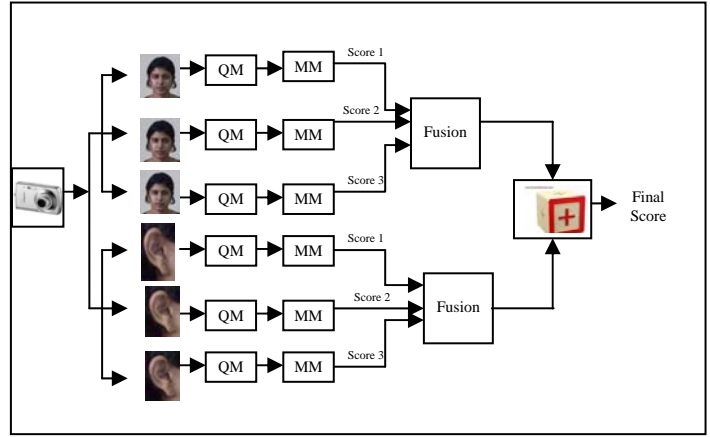

Figure 5. Multi-biometric fusion, QM: Quality Module, MM: Matching Module.

## V. EXPERIMENTAL RESULTS

The test of the proposed biometric recognition system consists in the evaluation of the quality modules, matching modules and the fusion block represented in Fig 5. The matching algorithms generate a score for each template comparison based on the distance between the tested and stored feature vectors. The Euclidean distance metric is used, as it achieves good results at a low computation cost [24]. The lowest distance score value indicates the best match.

The performance of individual biometric is shown in Fig 6 and Fig 7 below:

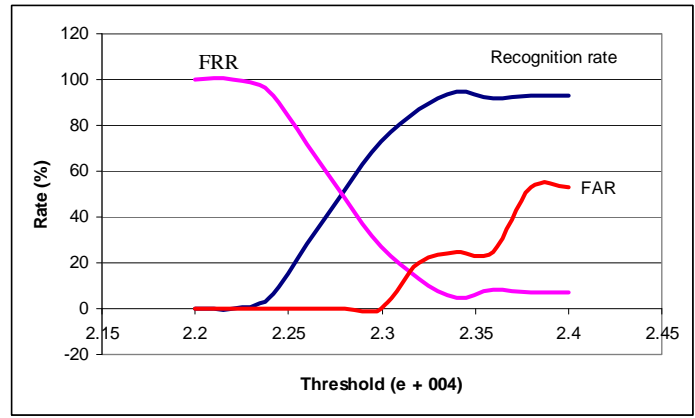

Figure 6. Face Recognition Performance Measures

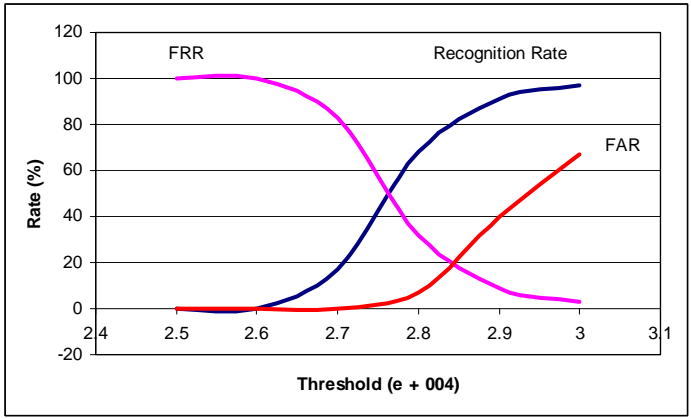

Figure 7. Ear Recognition Performance Measures

Using threshold values that maximize the correct recognition rates for each individual biometric, after fusion a FAR of 0 % was obtained, as illustrated in Table 1.

TABLE I. RESULTS FOR THRESHOLDS EQUIVALENT TO MAXIMUM CORRECT AUTHENTICATIONS

|  | Face | Ear | Multimodal Fusion |
|---|---|---|---|
| Recognition Rate | 94.7 % | 90.7 % | 96 % |
| FAR | 25 % | 40% | 0 % |
| FRR | 5 % | 9.3 % | 4% |

The unimodal face and ear biometric gives recognition rate of 94 % and 90.7 % respectively. When fused, the multi-modal gives a recognition rate of 96 %, showing an improvement in the accuracy. Also, both the FAR and FRR have been considerably reduced, showing that the multi-modal system implemented is more robust.

## VI. CONCLUSION

This paper proposes a multimodal biometric recognition system that exploits two modalities, namely face and ear recognition. With multi-sampling and fusion at decision level, a recognition rate of 96 % was obtained. Currently, we are working to enhance the recognition rate under uncontrolled environment so that it can be applied to surveillance applications.

**Nazmeen Bibi Boodoo** has done her degree in Computer Science and Engineering at the University of Mauritius. She is currently an MPhil/ PhD student at the University of Mauritius, Reduit, in the Department of Computer Science and Engineering. Her Research areas include Biometric Security and Computer Vision.

**R. K. Subramanian** is a professor at the University of Mauritius, Reduit, in the Department of Computer Science and Engineering.